Brief description on the state of the art of some local optimization methods:

# Quantum annealing

**Quantum annealing (also known as *alloy, crystallization* or *tempering*) is analogous to simulated annealing but in substitution of *thermal activation* by *quantum tunneling*.**

The class of algorithmic methods for **quantum annealing** (dubbed: 'QA'), sometimes referred by the italian school as *Quantum Stochastic Optimization* ('QSO'), is a promising metaheuristic tool for solving **local search problems** in **multivariable optimization** contexts. These problems usually consist in finding the maximum or minimum for a **cost function** that comprises several **independent variables** and a large number of instances.

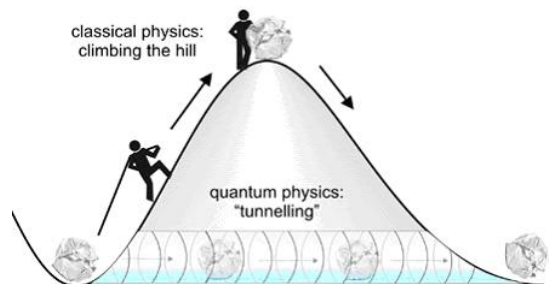

The **evaluation of cost** in this context must necessarily be computed in probabilistic terms, as given the large amplitude of the space of configurations (frequently *hamiltonian* matrices with the huge dimension of $2^N$ rows), the most common case is that an explicit, exhaustive evaluation of them all can not be performed because they are excessively numerous to be calculated in a reasonably practical time interval.

Let's think that, for a glass network with only five nodes, a $2^5*2^5 \sim= 1000$ elements matrix would need to be operated upon; with only ten nodes, this is boosted to over one million elements, and with sixteen nodes, to a whooping 4,300,000,000 (we dub this "**the curse of dimensionality**" or "*Hughes* effect", due to *Bellman*). A single **configuration** is thus defined as a '*tuple*' (or '*array*') of values over the whole set of independent variables. The value of the cost function depends on the configurations, being the **solution to the problem** set as the definite optimal configuration which minimizes, or maximizes, the cost function with some arbitrarily chosen confidence level or probability.

## ORIGINS AND NATURAL MODEL

In some way, all methods for annealing, alloy, tempering or crystallization are a **metaphor of nature that tries to imitate the way in which the molecules of a metal do order when magnetization occurs, or of a crystal during the phase transition** that happens for instance, when water freezes or silicon dioxide crystallizes after having been previously heated up enough to break its chemical bonds. If the cooling is slow enough ('*tempering*'), then the crystal generated this way will usually exhibit less imperfections (this is to say, it will be found in a **lower energy metastate**) than if it is frozen too fast (**higher energy metastate**). This physical model of nature is based upon the trend to minimize its free energy (in a *Helmholtz* sense) of any **ergodic** system, such as a closed thermodynamical system where every configurational state is equally likely to be reached.

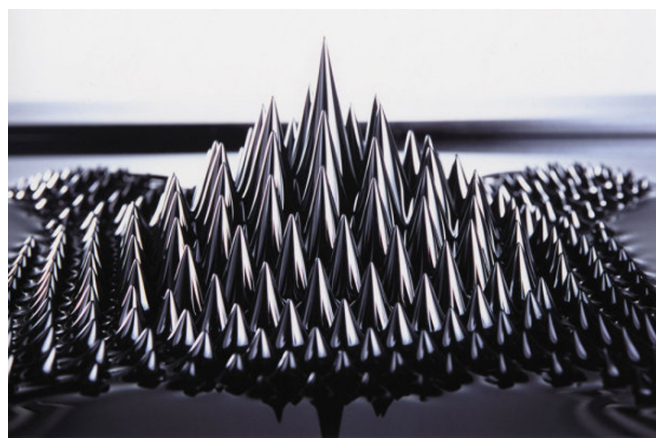

All methods for annealing are typically based on the *Monte Carlo algorithm*, which in order to generate sample states, repeats a **large amount of random sampling** on a hypercube of dimension 'N' (**space of solutions for the problem**), thus allowing to greatly reduce the computational complexity in trade of losing some statistical precision.

We will look further into the model for **ferromagnetic materials**, in which all of the charges are oriented in some determinate direction of space (called *spin*).

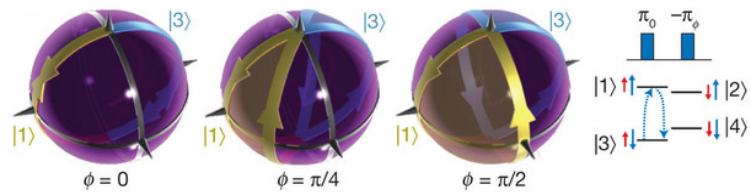

The **spin** in itself is really a typical representation of an angular momentum and does not literally imply for a particle to revolve around its axis.

The concept of "**crystallization**" or "**alloy**" gets intuitively clarified if we observe the appearance of imperfections in:

- freezing of ice blocks more or less translucent
- forging of mechanical traction resistant metal for knives, grinding gears or pressure vessels for industrial boiling
- industrial processes for fabrication of toughened glass in car windshields of for optical lenses without optical opacities in astronomy

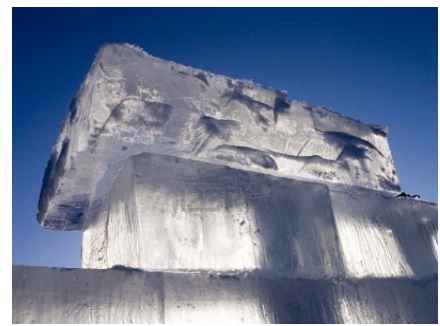

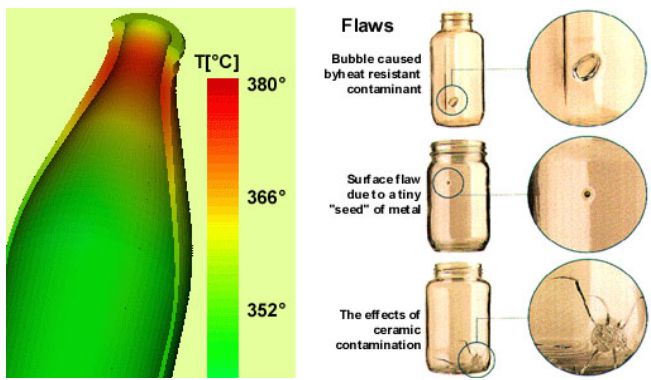

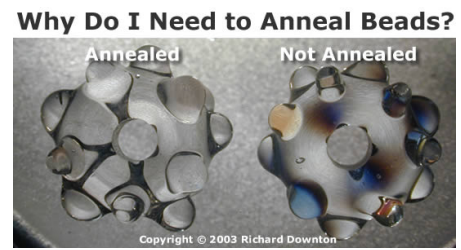

It may happen that the configuration space comprises many "large" thermal wells constraining the thermal relaxation along the time dimension. One **classic example for illustrating the criticality of a sufficiently slow crystallization** is what happens at a geological scale, deep within the lithospheric continental crust. When these large masses of subsurface rock are stressed under a geothermic gradient in conditions of pressure and temperature that are adequately sustained over time, they may crystallize orderly producing diamonds and *corunda* (such as ruby or sapphire), though when this process is too fast, only gemologically worthless pieces or coal and rubble will be generated.

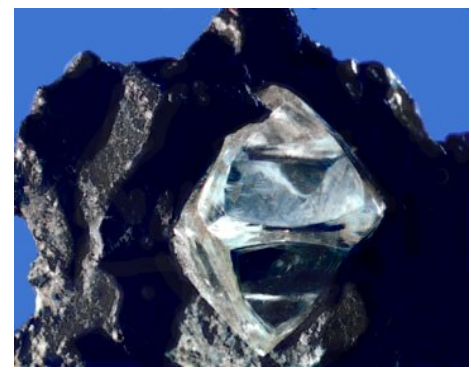

In the manufacturing industry, a similar process incorrectly executed will produce ices, glasses and metals which are fragile and have opacities, impurities and/or undesired changes of density. On the other hand, a finely executed and perfectly controlled process at the industrial level allows to create *graphenes*, carbon nanotubes, *fullerenes* and other carbon allotropes that are extremely useful given the excellent physicochemical properties that derive from purity in their crystallographic configuration.

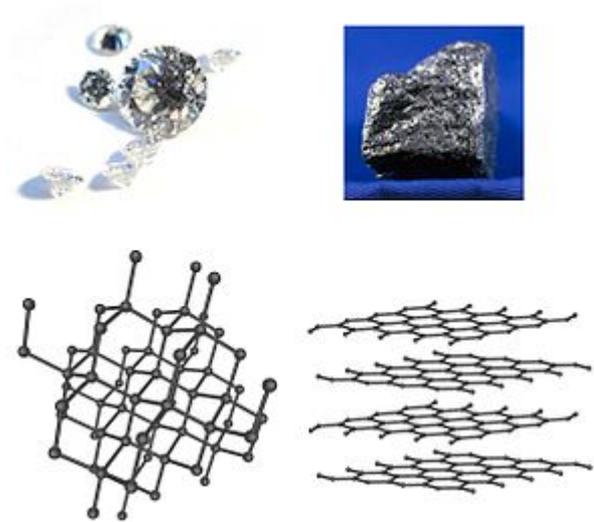

Among many other applications, the study of ferromagnets bears great relevance in the fabrication and operation of memories and hard disk drives for personal computers. In general, on these cases the temperature of the ferromagnetic material is rised over the *Néel* point in order to destroy its magnetic ordering, turning the substrate paramagnetic, and then being slowly tempered so that all its charges get oriented in a defined position that can later be read by an eraser head.

There exist a panoply of similar examples in nature, but let's see where the difference is between QA and other algorithmic annealing methods.

# INSIDE *QUANTUM ANNEALING*

QA is an algorithm class, similar to "*Simulated Annealing*" ('SA') from *Kirkpatrick* and others, that consists of an adaptation of the classical *Metropolis-Hastings algorithm*. Nonetheless, **QA uses a quantum field instead of a thermal gradient.** In order to explore the optimization problem's landscape, SA and its variants (such as *Parallel Tempering*) take advantage of **"thermal" fluctuations** associated to temperature gradients, while QA uses **"quantum" fluctuations** in this task.

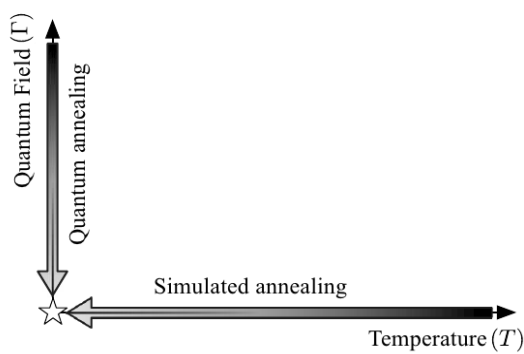

Fig. 1. Schematic picture of simulated annealing and quantum annealing. The star denotes the position of $T = 0$ and $\Gamma = 0$ which is the target position.

**A quantum fluctuation is a change in the amount of energy at a point in space for extremely short time lapses**, as a result of the **uncertainty principle** enunciated by *Heisemberg* (*cf. infra*).

In the natural metaphor upon which **metastates** are based, they may correspond to those of a **thermodynamical system**, that is to say, the one within which variations of temperature do exist, or either to metastates of a **quantum system**. The latter is the case for the wave function of the only electron that orbits an atom of the simplest chemical element: hydrogen (see picture).

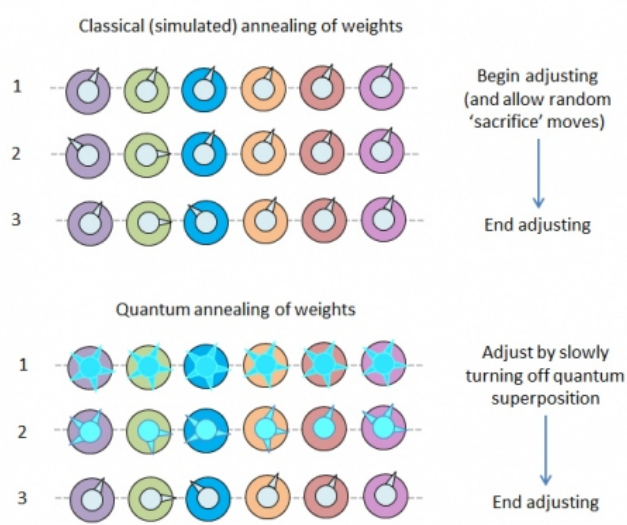

Within atomic hydrogen, the single-proton nucleus is orbited by an electron in a configuration manifold (a range of probability distributions) lacking the possibility of its precise location being determined with absolute precision. These effects appear at the subatomic scale, where quantum physical effects are measurable, though they also are observable with the help of experiments such as the **double slit** (*cf. infra*).

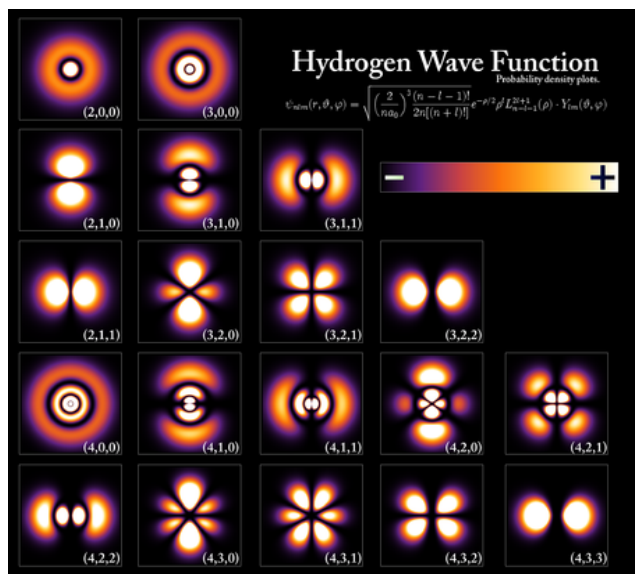

When the QA algorithmic class is applied to a minimization problem, a "**current state**" or candidate solution is replaced by a "**neighbor state**" chosen randomly (or else chosen by a more sophisticated method, analogous to "*adaptive SA*"), metastate which will have less energy in the **value of the target function**.

Alfonso de la Fuente Ruiz – 2011 http://www.linkedin.com/in/alfonsofr

When talking about minimization of a target (or maximization, which is the reciprocal problem), the fundamental idea is helping the system to escape the local minima thanks to the "**tunnel effect**" (*quantum tunneling*), that allows to take advantage of **wave-matter duality** (*cf. infra*) in order to traverse through the "solid" interstate potential barriers by **quantum jumps**, instead of trying to overcome those by **thermal jumps**.

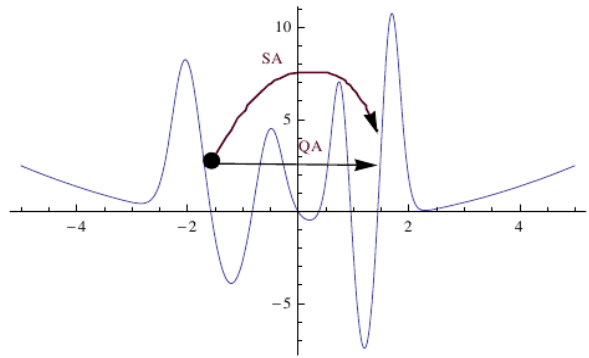

Figure 1: Thermal jumps, which in SA allow the exploration of the solution space, are substituted, in QA, by quantum jumps (tunneling).

The key controlling the metastate selection process is the **quantum tunneling width** or **quantum field strength**, a parameter determining the radius for neighbor states to be explored. The tunneling radius starts being very wide, so that **at first, the neighborhood comprises all of the search space** and as time goes by, it is gradually reduced during the computational simulation until the neighborhood shrinks enough that it barely differs from the current state (noise and error margins aside) instant that we could dub "**quantum collapse**" when we reach what we could call a "**quantum coherent state**" (see function of *Wigner* on picture).

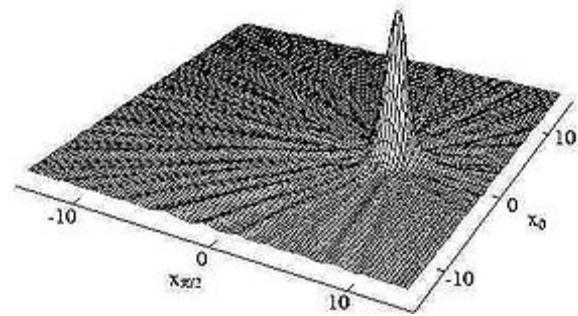

An adequate time limit should be chosen: **too fast a tempering** will keep the wave function from collapsing correctly, because when energetic conditions vary suddenly, crystallization (phase transition) will not correctly occur, as not enough time will be given for adaptation of the functional shape of the metastate so that it transforms its associated probability distribution. This capacity to adapt or resist sudden changes ('heat transfers' in thermodynamical contexts) is related to **adiabaticity**.

In natural quantum contexts for annealing, instead of **heat** transfers we will intuitively refer to changes in **entropy**, or disorder, of a system. One should bear in mind that even when cooling down to absolute zero temperatures some matter (conformed as a exemplary glass network known as "*spin ice*"), its residual entropy may still be quite noticeable. This effect occurs because of the so called "**zero point energy**" postulated by *Einstein and Stern*, meaning "the lowest energy that a physical mechanic-quantic system can possess", and it is the energy of the fundamental state (or "**ground state**") of the system. For example, liquid hydrogen does not freeze at absolute zero precisely because of its zero point energy.

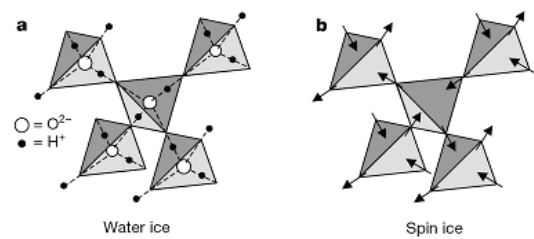

For these reasons it becomes clearly convenient that the quantum annealing process should be performed during as long a time span as possible, if a successful solution for the optimization problem is to be reached.

However, **too slow a tempering** will generally have the downside of a greater cost in terms of sheer computing power, work memory and processing time, that are main hindrances in high-dimensional configuration spaces with a large number of instances. Be viewed that some studies even refer glass networks of infinite range. It is also convenient to adjust the **optimal equilibrium** between the quality of the annealing process and the computational power to be used, among a range of different algorithmic variations for QA (cf. infra).

If the **tempering is too fast**, it is possible that some interstate energy barriers are to be found within the configuration space that, being too wide, will avoid tunneling through by quantum jumps, thus leading the algorithm to get "stuck" at local minima (see the next picture, referring to a problem of *Variacional Bayes inference* - 'VB' solved by QA).

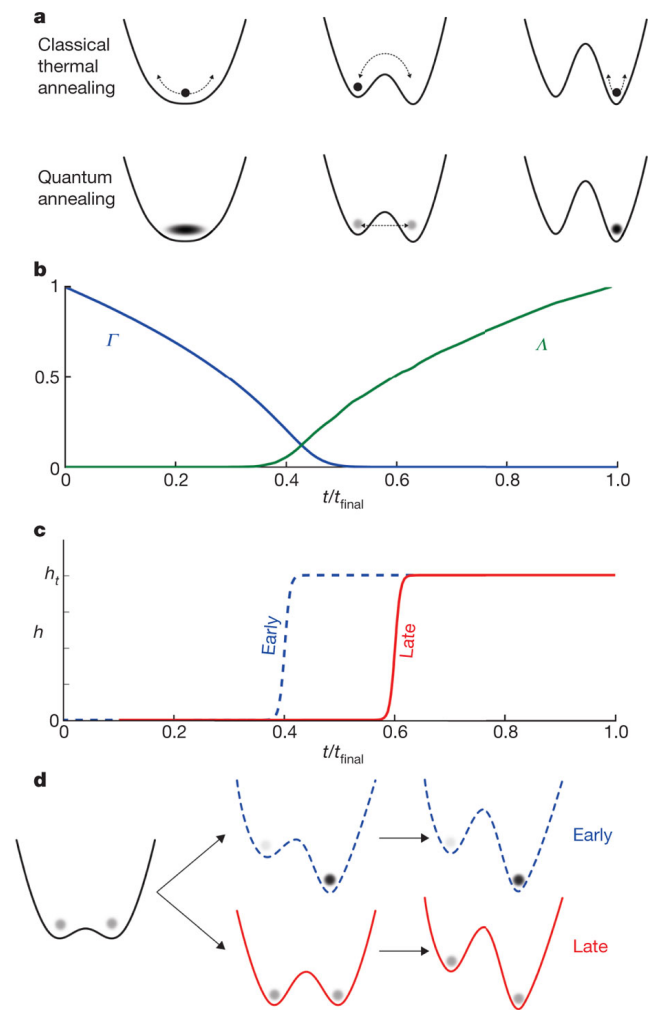

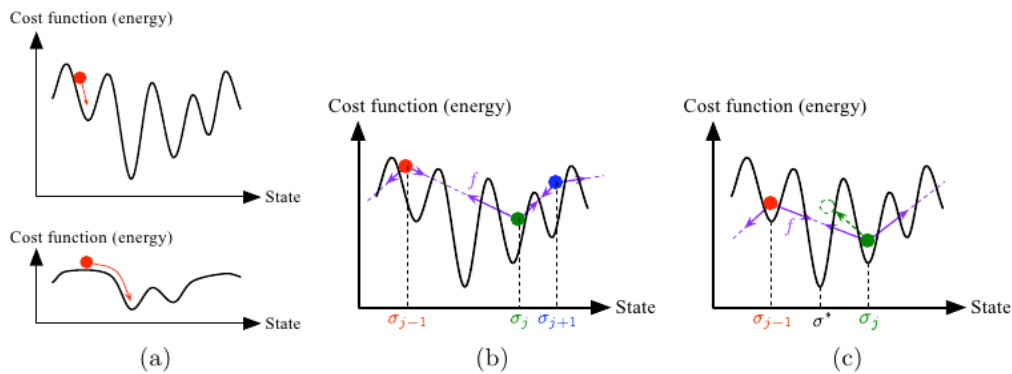

Figure 1: (a) Schematic picture of SAVB. (Upper panel) At low temperature, the state often falls into local optima. (Bottom panel) At high temperature, since the energy landscape becomes flat, the state can change over a wide range. (b) and (c) Schematic picture of QAVB. (b) QAVB connects neighboring SAVBs. (c) $\sigma_j$ can reach $\sigma^*$ owing to the interaction $f$. It seems to go through energy barrier.

In the context of quantum annealing, it is of major concern to know **how does residual energy decrease along with the simulation**, or equally: to study the **adiabaticity** of a system that evolves quantically. In order to understand this model entirely, we should first dedicate some words to briefly explain the concepts that this particle physics natural metaphor incorporates.

**Wave-matter duality** is a concept that was developed by *De Broglie* in his 1924 PhD thesis, and previously introduced some centuries before thet date by *Huygens* and some other authors, in order to explain **reflection** and **diffraction** of light, a phenomenon chromatically observable, for instance, when looking at the rainbow or the surface of a CD-ROM.

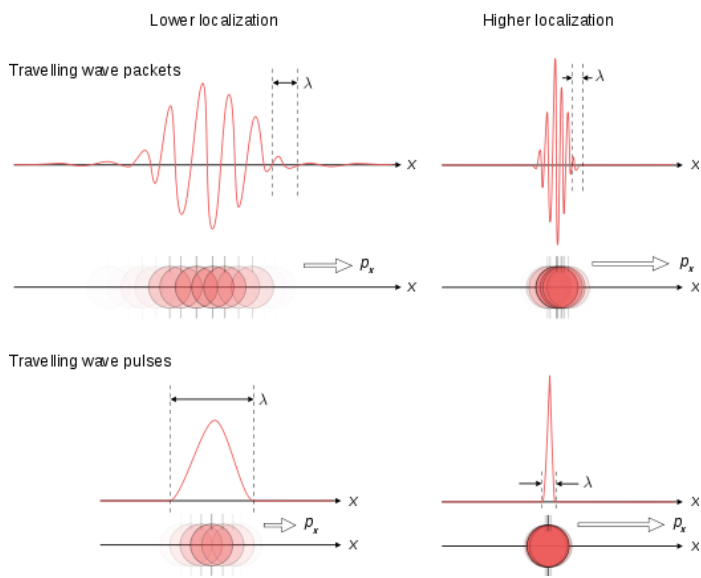

In the XIX century, the **double-slit experiment** from *Young* gave as a result some **interference patterns** that came to demonstrate the **wave theory of light**, and thus, that matter and energy could either behave like waves of particles under different observational circumstances. His experiment suggested that a subatomic particle was capable of getting through both slits -simultaneously- and even to interfere with itself, as if it was a wave. This wave-particle duality quickly became a fundamental concept in quantum mechanics, that went far beyond *newtonian* classical physics.

Shortly after *De Broglie*, in 1927, *Heisenberg* enunciated his **uncertainty principle** establishing that we can not know simultaneously and with arbitrary precision some pairs of physical variables, such as the position of a particle and its linear momentum. In fact, the product of both imprecisions would in the best-case scenario be equal to the minuscule *Planck* length.

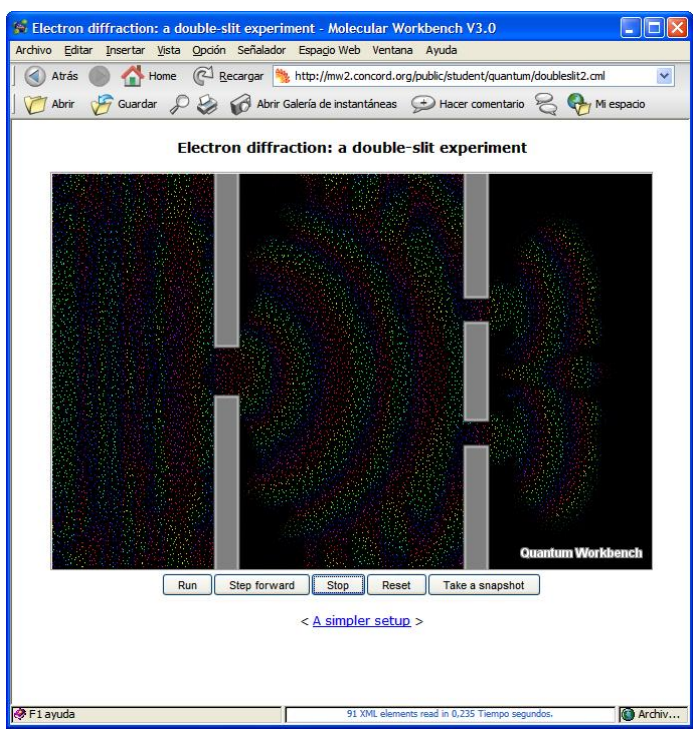

In other words: something can not be observed as a wave and as a particle at the same time. This also implies the strange corollary that the mere observation of an experiment modifies its outcome, because for it to be evaluated it is necessary to intervene in its **wave function collapse**.

The impossibility of observing this process by experimental means is what in quantum mechanics is called the "**measurement problem**".

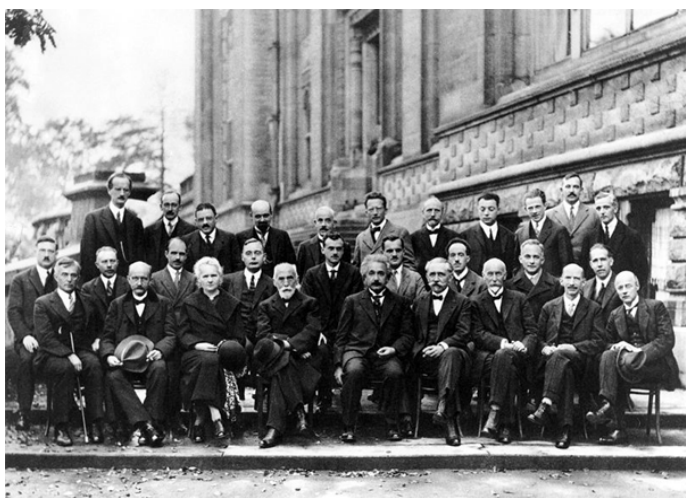

Even though it already arose with the studies on radioactivity in previous decades, it was the very same year of 1927 when *Friedrich Hund*, professor at Göttingen, discovered the "**quantum tunneling effect**", that seemed to be an incredible way, derived from quantum mechanics equations, of passing through supposedly impassable walls (or more precisely: **energy potential barriers**).

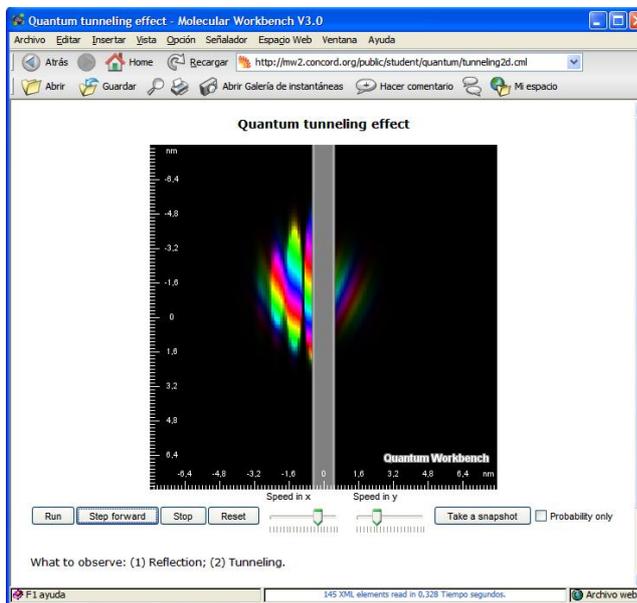

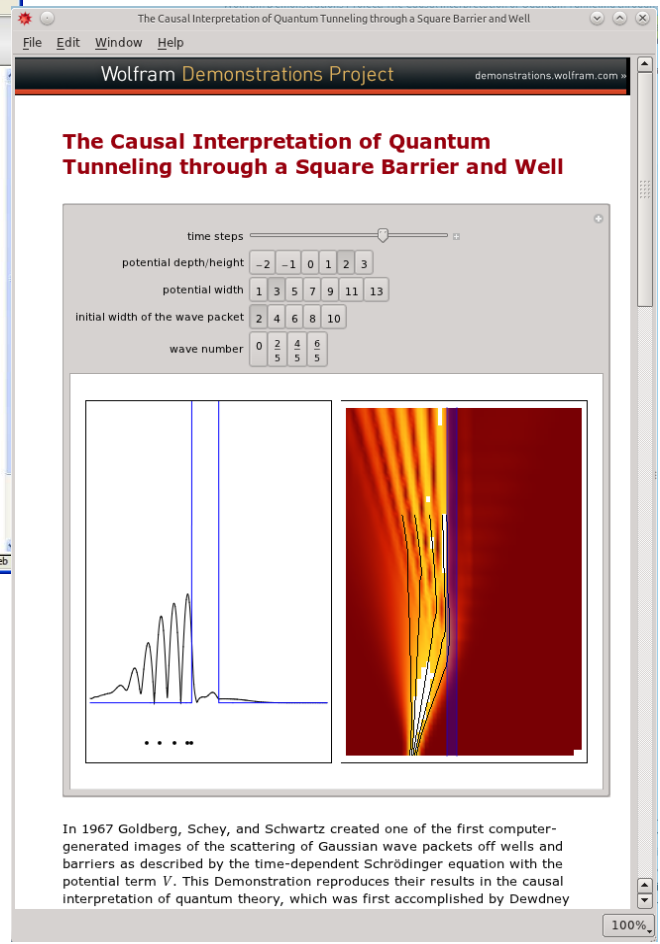

This phenomenon has been observed, studied and controlled, being well known in the scope of quantum physics (see pictures and Java *applets* from *Concord*, as well as simulations from *Wolfram Mathematica*), and it's the base for designing the *Quantum Annealing* algorithm class. The idea is that, as the wave function is solved, *Heisemberg* uncertainty principle associates it to a definite **probability density**, related to the place that a particle occupies at any instant in time. The implication that **no solutions do exist with an exact probability of 0 (or 1)** means that the particles will have a certain probability of getting out of a potential well, that will be lower as shallower and thinner the barrier is, but never null. From there arose the analogy that the particle seems to excavate a **quantum tunnel** through a potential barrier in order to escape the well.

In essence and following the authors that have worked on its implementation, QA uses a **classical function of energy cost** that is added to a **quantum-kinetic term traveling over the problem space** as a function of time. At the beginning of the algorithm, the kinetic term is very large, and it is then reduced tending towards zero, down to a precise time limit. Working that way, the quantum state of a system, that initially is a fundamental quantum state, evolves with time according to the *Schrödinger* and *Fokker-Planck* **wave equations** until a final solution state is reached.

$$\hat{H}\left|\Psi(t)\right\rangle = i\hbar \frac{d}{dt}\left|\Psi(t)\right\rangle = \frac{\hat{\vec{\mathbf{P}}}^2}{2m}\left|\Psi(t)\right\rangle + V(\hat{\vec{\mathbf{r}}},t)\left|\Psi(t)\right\rangle$$

The previous equation expressed in angular parenthesis notation ("*bra-kets*", introduced by *Dirac* in order to describe quantum states) determines the probability that a **wave's quantum superposition state collapses** in a simple quantum state (represented by a vector).

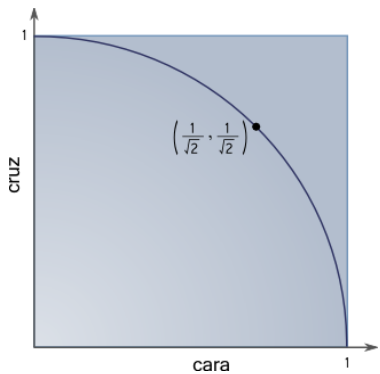

Let's see a simple example: We have a coin. When tossing it will be considered a wave that can end in two possible distinct states: if we write in *Dirac*'s notation, those would be |head⟩ and |tails⟩ which will be **orthogonal** (perpendicular) by definition. In classical physics we will say that a coin must be in one of these two states at all times. However in quantum physics, if those two states are possible, also will be the combinations of them which lie at a distance of one unit from the origin of coordinates. What we basically say is that a "quantum coin" will be able to stay in both states at the same time, heads and tails, with the same proportion (half heads or '*cara*' and half tails or '*cruz*', see picture) or a different proportion. The **superposition state** of head and tails will be maintained only when the coin "is not observed", because when its function is tested, the step is taken from the superposition state to be evaluated either in |head⟩ or |tails⟩ with a 50% probability each (or a different proportional mixture).

If instead of a coin we rolled a cubic die, we would act similarly, but within a space of dimension six instead of two. For a subatomic particle, we would use **Hilbert spaces** that allow us to extend the techniques of *euclidean* space to any arbitrary dimension, even infinite.

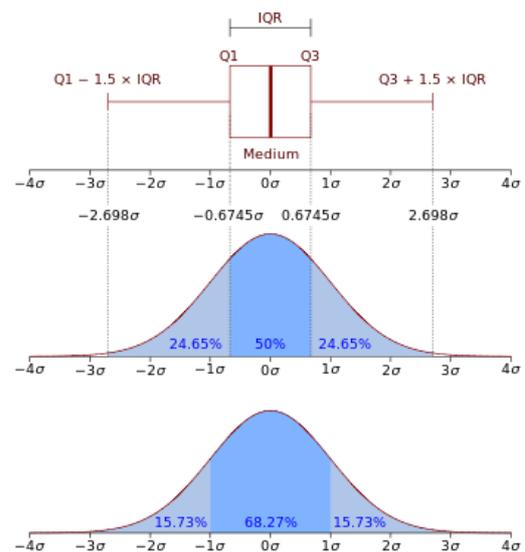

Interpreted this way (in no-relativistic contexts), **time-dependent *Schrödinger*'s wave equation** corresponds to a **probability amplitude** (see picture), specified like a number in the complex plane, that we will be able to algorithmically tune so that it falls within arbitrarily small margins, in conformity with the required precision.

These margins will describe the exactitude of the **quantum condensate** thus collapsed in probabilistic terms, giving a **point solution to the optimization problem over its configurational space**.

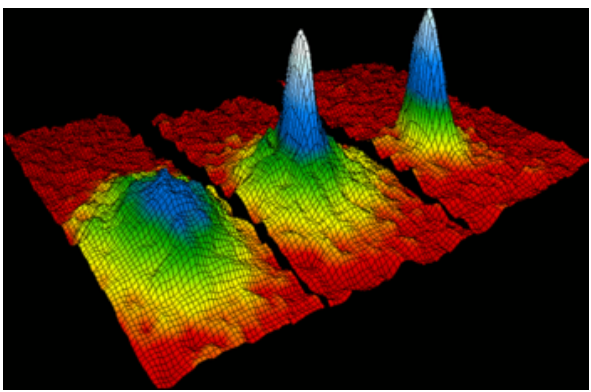

On its side, the *Fokker-Planck* equation (also known as *Kolmogorov forward equation*) is a non-reversible equation representing the evolution in time of a probability distribution (see programmatic examples at *Wolfram*). Its main advantage is that it allows the calculation of a wave thanks to a certain **drift potential** and to a **diffusion constant**:

$$D_{\text{eff}}(T) = \gamma a^2 e^{-B/k_B T}$$

According to the indian school (*Chakrabarti* et al.), we must study the physical underlying process for the phase transition, quantum-wise. In order to fully understand QA we must think in terms of "**friction**", "**viscosity**" and "**entropy**", because they will be the ones that determine the quantum tunneling process and thus the algorithm's **loop and stop criteria**.

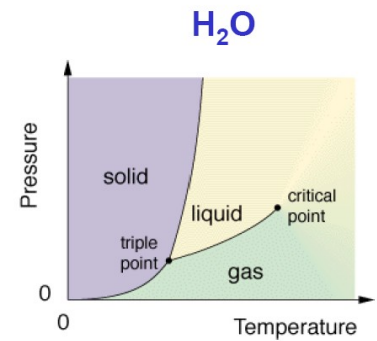

As it is well known, similar cost functions may be extracted from using *Boltzmann* constant in the thermodynamical processes which are modelled by the SA (*simulated annealing*) class.

The quantum physics in which QA is based upon considers tiny "***quanta***" or "individual acting packets" that work in a discrete manner, in opposition to the idea of a uniform ***continuum***. The author *Kovtun* and others, conjectured that the rate between viscosity and entropic density (or "disorder" density of a system) is bounded by *Dirac* constant ($\hbar = h/2\pi$ also called "reduced *Planck* constant"). On another side, generalizing QA algorithmics to any possible configuration space, maybe we should not impose a similar physical limit, but to abort simulation once the required precision is reached.

In order to grasp a glimpse of the **extreme dimensions of this limit**, the environment of the *Planck* distance moves around powers higher than the $35^{th}$, meaning a 1 followed by 35 (or maybe more) zeroes, which seems enough for manageable problems on current supercomputers that move around the order of *petaflops* (10 to the $15^{th}$ power floating point operations per second) which furthermore have evolved similarly in terms of memory capacity (in the order of *terabytes* of RAM and *petabytes* of storage disks, respectively). Nonetheless, it must be borne in mind that **the quantity of information that is needed to represent a quantum state grows exponentially with the size of the problem** in a classical computer. So to avoid managing the immense resulting matrices, we usually perform **Monte Carlo simulations** (cf. infra) that complement the wave equations.

# PRACTICAL APPLICATIONS

The QA algorithmic class has demonstrated being more efficient than its classical counterpart (SA) in various and diverse applications, such as:

- **Combinatorial optimization**, like the classical "*Traveling Salesman Problem*" ('TSP') and other *NP-hard problems*. In the picture the tests performed by *Tadashi Kadowaki* in his thesis are reproduced, clearly showing that the likelihood of finding the minimal path over a graph (for certain field values) is considerably better when using QA instead of SA.

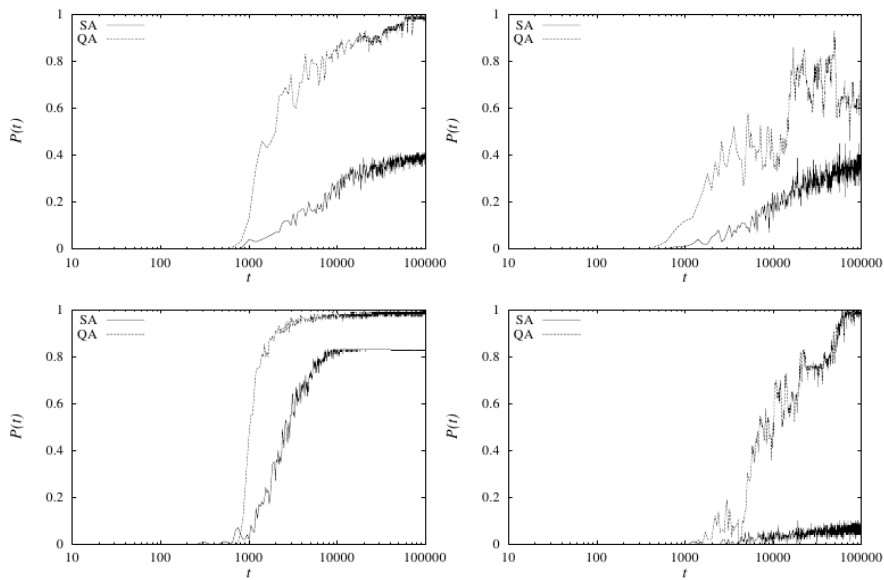

Figure 4.7: Probability to find the minimum-length. The scheduling is $10/\sqrt{t}$.

- Within the previous class of optimization problems, **Ernst Ising ferromagnetic model for disoriented magnets** bears special relevance, being a simplified model for crystallization used interdisciplinarily. It directly relates to the *spin glasses* than result in ferromagnets upon orientation (see picture).
- **Integer factorization**, an essential task in cryptography (see *Shor algorithm*) and consequently in secure telemetric data transmission.
- **Search in unsorted databases**, including web search engines on-line: important computer science problems with wide practical applicability nowadays (see *Grover's algorithm* and picture), that can be modeled with **quantum walks**.

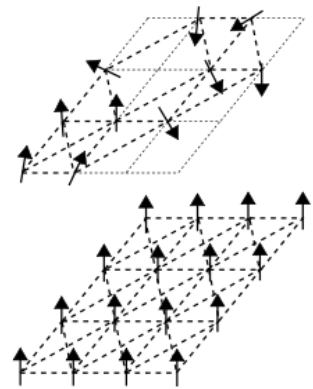
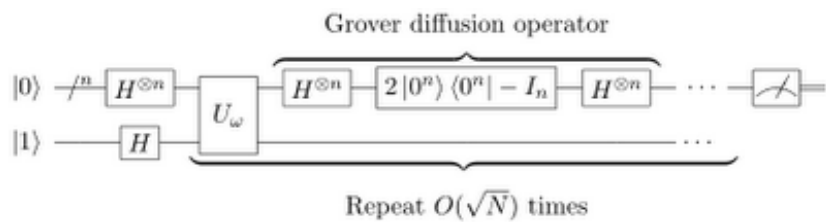

- **Pattern recognition** and other *NP-hard problems* that appear in heuristic learning for tasks in *Artificial Intelligence*. Among them: **Fuzzy image reconstruction** and object recognition in correlated images (see pictures, cf. *Inoue & Chakrabarti*)

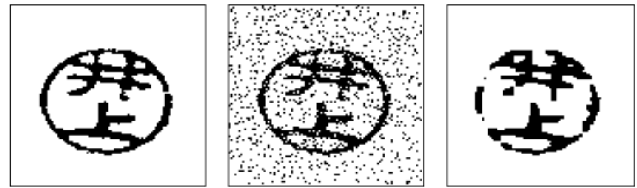

Fig. 1. A typical example of image data retrieval. From the left to the right, the original $\{\xi\}$, degraded $\{\tau\}$, and the recovering $\{\sigma\}$ images. The above restored image was obtained by quantum annealing. The detailed account of this method will be explained and discussed in last part of this chapter

- **Protein folding problems**, that are physiochemical simulations in which atoms and molecules are placed in specific, univocal positions, similar to those of a crystal (see pictures below). The main difference of a protein chain and a crystal is that the arrangement of a protein's atoms is not periodically ordered, in spite of which, proteins may be classified as **ordered structures** because they present different spatial configurations where attraction and repulsion forces act, determining their final shape upon folding along a kinetic evolution in time.

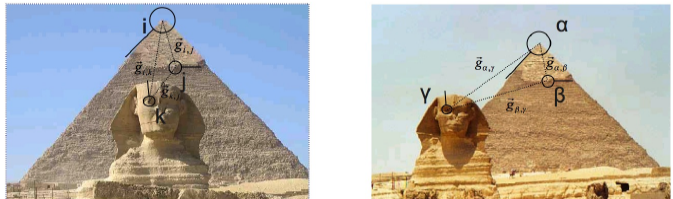

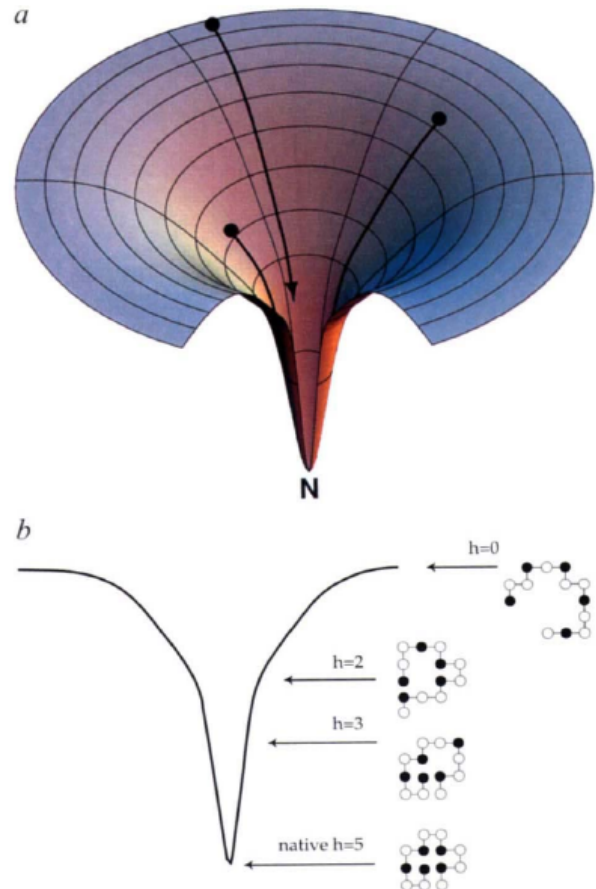

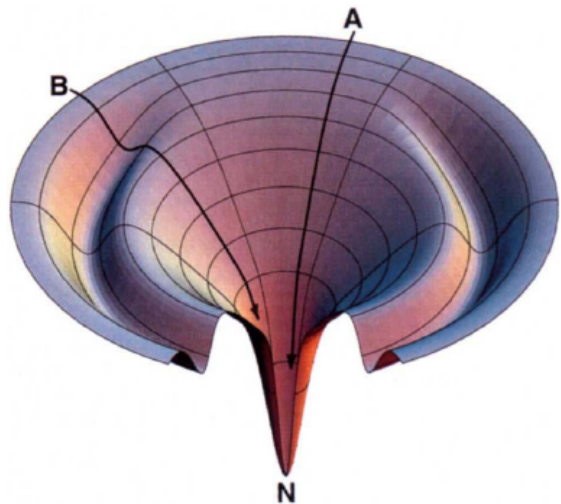

Fig. 5 Moat Landscape, to illustrate how a protein could have a fast-folding throughway process (A), in parallel with a slow-folding process (B) involving a kinetic trap.

nature structural biology • volume 4 number 1 • january 1997

Fig. 3 *a*, An idealized funnel landscape. As the chain forms increasing numbers of intrachain contacts, and lowers its internal free energy, its conformational freedom is also reduced. *b*, A slice through (a). In the lattice model, black beads represent hydrophobic monomers and white beads represent polar monomers, $h$ is the number of hydrophobic contacts. Exact enumeration studies show that there are many open conformations, fewer compact conformations, and only one conformation having $h=5$. An ensemble of molecules can reach the global minimum of free energy (satisfying Anfinsen's thermodynamic hypothesis), and do so quickly (satisfying Levinthal's concern), even though each chain follows its own route, not a single pathway.

# IMPLEMENTATIONS AND VARIANTS

For its implementation as computer software, the QA algorithm class may be developed stemming from a range of stochastic methods such as **Monte Carlo**, of demonstrated *convergence* (cf. *Satoshi Morita & Hidetoshi Nishimori*).

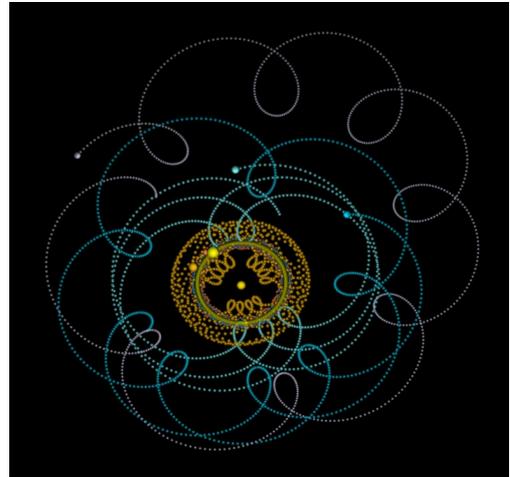

The Monte Carlo methods (dubbed: 'MC') are tipically used to simulate, among others, the **'N' body problem**. In this setting, three or more particles interact gravitationally with one another in free fall, quickly drifting to hardly predictable **chaotic orbits** (see picture), as their computation requires the solution of many multidimensional integrals.

Diverse variants of QA have been designed, such as a **hybrid kind** incorporating both thermal and quantum fluctuations, modifying quantum field intensity as metastable states are being found.

Within the quantum algorithms class, one of the most heavily documented basis for QA, is *Adiabatic Quantum Computation* (*'AQC'*), that should not be confused with properly defined quantum annealing, because it is more generic. In AQC, the **ground state** is adiabatically deformed until a solution state is reached. Nonetheless, it has been shown that AQC fails for *NP-complete problems* and presents some other inconveniences.

A colorfully detailed explanation of this scheme can be found in *Ilievski*, *Dwave* corporation and a videotalk on the subject at *Google Talks*.

Generally speaking, on a **classical computer architecture** the quantum states (in the so-called *qubits*) will be emulated with conventional binary states, so the computational performance will be subject to an onerous exponential factor.

Even so, there already exist some hardware-built rudimentary **quantum experimental microprocessors** that some corporations such as *D-Wave* are developing, for *Lockheed-Martin* aerospatial industry (see picture), presumably capable of cutting down algorithmic complexity in a radical manner, although current prototypes do not integrate a large amount of *qubits* as of today.

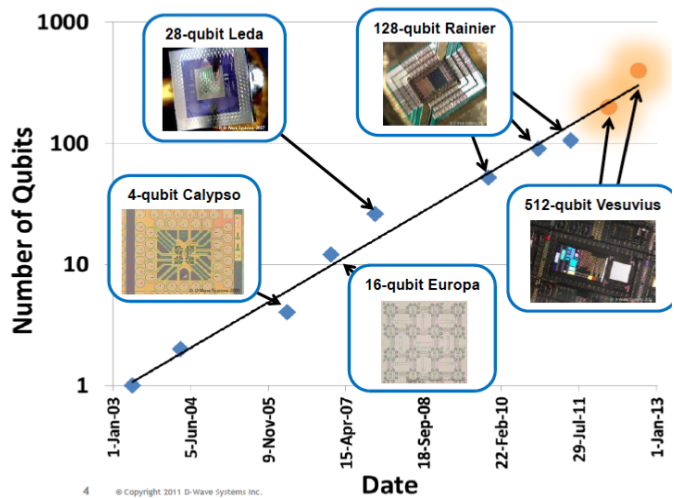

Nonetheless it should be advised that its relative credibility has not yet fully convinced some sectors of the scientific community.

**Source code** may be semi-automatically generated with the Java *applet QusAnn* (see picture) for running QA on a quantum computer. However, being access to this kind of experimental computers currently very restricted, this can not easily be tested, though as it seems, it is possible to program those in *Python* anguage.

Returning to **QA on conventional computer architectures**, some differentiated **quantum fluctuations** have been designed too (see for instance: *Tanaka, Tamura, Sato & Kurihara*) and also ways for **iterative tuning of the quantum field, with restart**, as well as studies of **hybrid methods** applied to clusterization problems.

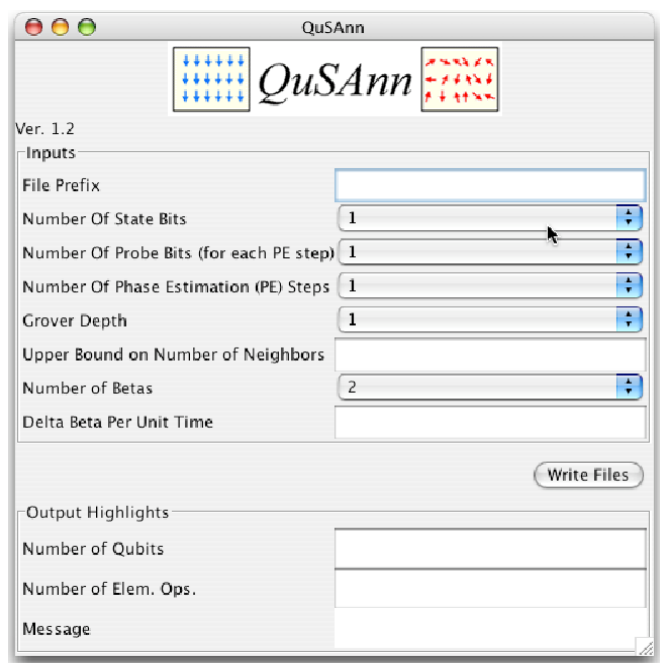

Figure 1: Control Panel of QuSAnn

On the other hand, the study of the **action radius for quantum interactions** in spin glasses has been very advanced (see *Takahashi, Nishimori and Martín-Mayor*).

The set of variants for the QA algorithm is what we name "QA algorithmic class". Some authors such as *Satoshi Morita* have been targeting the **asymptotical optimization of crystallization speed**, based upon the reduction of the **excitation probability** thanks to the **quantum adiabatic theorem**.

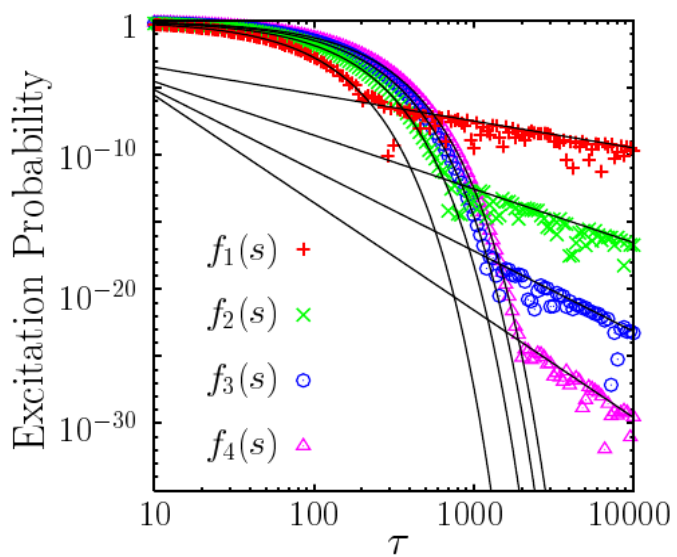

Most of the times when QA overcomes other methods in performance, it happens that the solutions landscape (configuration space) hosts **very thin barriers** that separate **very deep chasms between local minima**. Such is the of a parametrized *Rastrigin* function (see pictures) and its variants for spaces of different dimensionality.

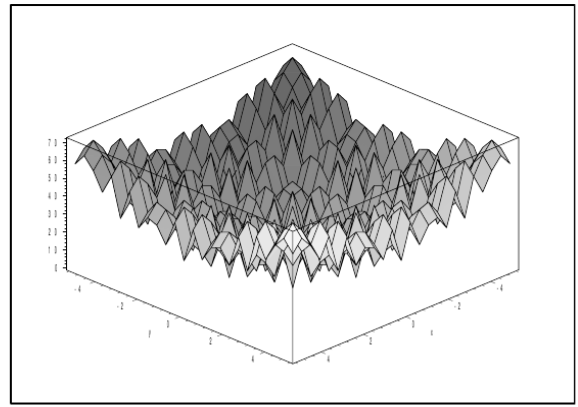

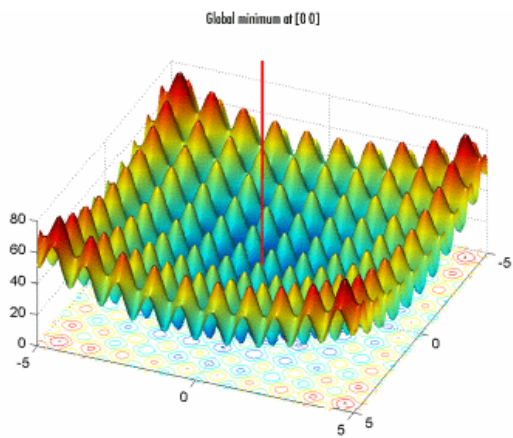

Rysunek 5: An overview of Rastrigin's function in 2D, $f(x,y) = 10 \cdot 2 + [x^2 - 10\cos(2\pi x)] + [y^2 - 10\cos(2\pi y)]$

### 2.5 Rastrigin's function

Rastrigin's function is based on the function of De Jong with the addition of cosine modulation in order to produce frequent local minima. Thus, the test function is highly multimodal. However, the location of the minima are regularly distributed. Function has the following definition

$$f(x) = 10n + \sum_{i=1}^{n} \left[ x_i^2 - 10\cos(2\pi x_i) \right]. \quad (5)$$

Test area is usually restricted to hyphercube $-5.12 \leq x_i \leq 5.12$, $i = 1, \ldots, n$. Its global minimum equal $f(x) = 0$ is obtainable for $x_i = 0$, $i = 1, \ldots, n$.

In these environments. It is very likely that SA (classical tempering) gets stuck at a local minimum, given that the potential barriers opposing the **thermal jump** are very high and at the same time numerous. Because the barriers are very thin, the **quantum jump** will have a higher success rate, predictably rendering QA much more efficient than its counterpart.

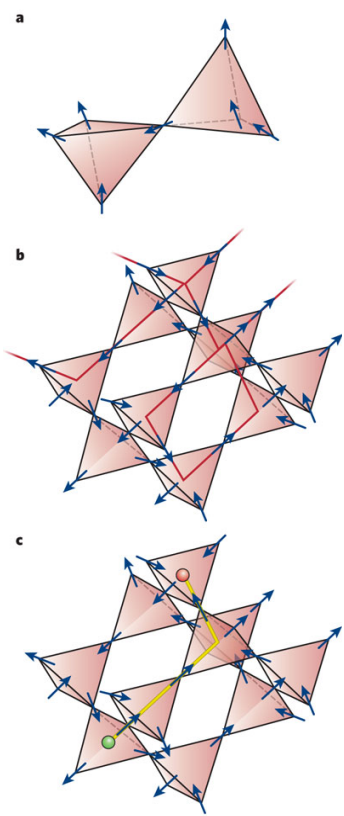

The (thermal) energy barriers in the configuration space can have a quasi-infinitum size, scope that we call a *kinetically constrained blocking*, and may be imagined as the presence of impurities in a magnet, a glass or some crystalline molecular network of other kind (all of them modeled like "*spin glasses*" with *Bravais 'lattice' crystalline networks*). These impurities within a spin glass may naturally block or reflect the direction of the electromagnetic field, for instance distorting the former's magnetization or transparency properties.

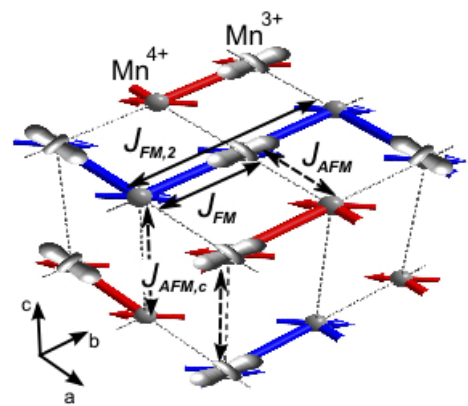

*Jörg* and other authors have shown under which circumstances **QA finds performance obstacles**: precisely in a first order transition where two states whose free energies cross are far away from one another over the phase space, implying the **existence of very wide and shallow potential energy barriers** (the inverse situation of that of *Rastrigin*).

Different programmatic approximations and **implementations of QA on classical computers** have been developed, some of which were studied by *Lorenzo Stella* in his thesis. Precisely, it may be interesting to see the scheme based on the *Path Integral Monte Carlo* ('**PIMC-Q[T]A**', see picture) for *Quantum [Thermal] Annealing*, and the one based on *George Green's function*('**GFMC-QA**'), also of the Monte Carlo kind and using diffusion equations. The **QuMax** package from the GNU Scientific Library ('GSL') contains some implementations of the Monte Carlo quantum algorithms, though they seem to be currently in clear disuse.

Associated to Barcelona Super-Computing Center, *Daniel Lecina Casas* has worked in QA algorithms for solving some **graph coloring problems**. To be seen next is the algorithm for this precise setting as suggested by *Alan Crispin* and *Olawale Titiloye*.

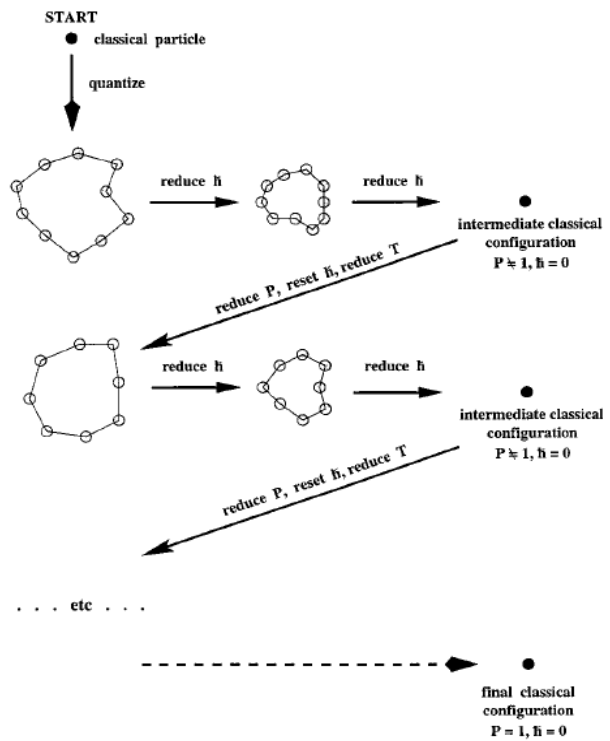

**Figure 3.** QTA-PIMC scheme used in this paper, illustrated with a single classical particle. The classical particle is quantized into $P$ Trotter beads (shaded circles). The quantized particle is annealed back to the classical regime in two ways: by decreasing $\hbar$, and by decreasing $P$. For a fixed $P$, the quantized particle is annealed by decreasing the value of $\hbar$. When $\hbar$ reaches 0, all the Trotter beads would converge back to a single point which corresponds to an intermediate classical configuration (solid circle). At this juncture, we remove one or more of the overlapping Trotter beads. We reset the value of $\hbar$ to $\hbar_i$, reduce $T$ by $dT$ (if thermal annealing is desired), and repeat the whole process again. The final classical configuration is obtained when both $P$ and $\hbar$ are annealed to the classical values of 1 and 0, respectively. This configuration would be the global minimum of the system, although often the global minimum would already have been found in one of the intermediate stages.

**Algorithm 2.** QA-col: Graph coloring with QA
1: **Input**: Graph $G$, number of colors $k$, the number of replicas $P$, $T_0$, $\Gamma_0$ and *MaxSteps*
2: **Output**: Best coloring configuration found
3: Initialize $T = T_0$, $\Gamma = \Gamma_0$ and a set of $P$ random coloring configurations $\{\omega_\rho\} := \varpi$
4: **repeat**
5:    randomly shuffle the order of replicas
6:    **for** $r = 1, \ldots, P$ **do**
7:      select replica $\rho$ in position $r$
8:      **repeat**
9:        Randomly select vertex $v$ from the list of vertices involved in conflicts in $\omega_\rho$
10:       Only in $\omega_\rho$, move $v$ to a new randomly selected color class, to derive $\omega'_\rho$ and hence $\varpi'$
11:       $\Delta \mathcal{H}_{pot} = \mathcal{H}_{pot}(\omega'_\rho) - \mathcal{H}_{pot}(\omega_\rho)$
12:       $\Delta \mathcal{H} = \mathcal{H}(\varpi') - \mathcal{H}(\varpi)$
13:       **if** $\Delta \mathcal{H}_{pot} < 0$ **or** $\Delta \mathcal{H} < 0$ **then**
14:         $\varpi = \varpi'$
15:       **else**
16:         With probability $\exp(-\Delta \mathcal{H}/T)$, **set** $\varpi = \varpi'$
17:      **until** iterations $= M.N$
18:    **end for**
19:    $\Gamma = \Gamma - (\Gamma_0/MaxSteps)$
20: **until** termination condition.

Aditionally, a pseudcode implementation of the QA algorithm is included below, according to *Diego de Falco* and *Dario Tamascelli*, together with its associated quantum transitions, all designed for conventional computer architectures.

**Procedure 1** Quantum annealing

**Input:** initial condition $init$; control parameter $\nu$; duration $t_{max}$; tunnel time $t_{drill}$; local opt. time $t_{loc}$.

$t \leftarrow 0$;
$\epsilon \leftarrow init$;
$v_{min} = cost(\epsilon)$;
**while** $t < t_{max}$ **do**
  $j \leftarrow 0$;
  **repeat**
    $i \leftarrow 0$;
    **repeat**
      $\epsilon \leftarrow$ Quantum Transition$(\epsilon, \nu, t_{max})$;
      **if** $cost(\epsilon) < v_{min}$ **then**
        $v_{min} \leftarrow cost(\epsilon)$;
        $i, j \leftarrow 0$;
      **else**
        $i \leftarrow i + 1$;
      **end if**
    **until** $i > t_{loc}$
    $epsilon \leftarrow$ Local Optimization$(\epsilon)$.
    **if** $cost(\epsilon) < v_{min}$ **then**
      $v_{min} \leftarrow cost(\epsilon)$;
      $j \leftarrow 0$;
    **end if**
  **until** $j < t_{drill}$
  draw a trajectory of length $\nu t_{max}$ and jump there.
  Local Optimization$(\epsilon)$
**end while**

**Procedure 2** Quantum Transitions

**Input:** initial condition $\epsilon$; chain length $\nu t$; set of neighbours to estimate $Neigh$

**for all** neighbour $k \in Neigh$ **do**
  estimate the wave function $\psi_\nu(k)$;
**end for**
$best \leftarrow$ select a neighbour in $Neigh$ with probability proportional to $\psi_\nu$
**return** $best$

# REFERENCE WORKS

Gathered in blocks depending on their nature, the appear approximately in the same order as their contribution to this article, within each block.

**Main references:**

- Optimization through quantum annealing: theory and some applications
  https://iris.ucl.ac.uk/research/browse/show-publication?pub_id=190257&source_id=3
- An introduction to quantum annealing, de Falco & Tamascelli, univ. Milano http://arxiv.org/abs/1107.0794
- An introduction to quantum annealing, Rose & Macready (Dwave)
  http://dwave.files.wordpress.com/2007/08/20070810_d-wave_quantum_annealing.pdf
- Studies of Classical and Quantum Annealing, thesis by Lorenzo Stella
  http://www.ted.com/talks/aaron_o_connell_making_sense_of_a_visible_quantum_object.html
- Study of Optimization Problems by Quantum Annealing, thesis by Tadashi Kadowaki
  http://cdsweb.cern.ch/record/550293?ln=es
- Quantum Annealing and related optimization methods (Springer), Chakrabarti et al.
  http://books.google.com/books/about/Quantum_annealing_and_related_optimizati.html?id=m04GZnNyJ7MC
- Mathematical foundation of Quantum Annealing, Satoshi Morita & Hidetoshi Nishimori
  http://arxiv.org/abs/0806.1859
- Convergence theorems for Quantum Annealing http://arxiv.org/abs/quant-ph/0608154

**Academic articles:**

- Quantum Annealing of Hard problems http://arxiv.org/abs/0910.5644
- Quantum Annealing of a Disordered Magnet, Gabriel Aeppli
  http://www.sciencemag.org/content/284/5415/779.abstract
- Quantum annealing of a disordered magnet – Brooke et al. http://cdsweb.cern.ch/record/499759?ln=es
- Quantum annealing of the random-field Ising model by transverse ferromagnetic interactions
  http://arxiv.org/abs/quant-ph/0702214
- Quantum annealing in the transverse Ising model http://arxiv.org/abs/cond-mat/9804280
- Quantum Annealing of the Traveling Salesman Problem http://cdsweb.cern.ch/record/711759?ln=es
- Microscopic Properties of Quantum Annealing -- Application to Fully Frustrated Ising Systems - Shu Tanaka
  http://arxiv.org/abs/1106.0555
- Quantum Annealing and the Schrödinger-Langevin-Kostin equation, de Falco & Tamascelli, univ. Milano.
  http://arxiv.org/abs/0812.0694
- Phase transitions and the perfectness of fluids – Jiunn-Wei Chen, Nat. Taiwan univ.
  http://arxiv.org/abs/0709.3434 http://cts.phys.ntu.edu.tw/cts/download/20081216_(Jiunn-WeiChen).ppt
- Scalable architechture for quantum adiabatic computing of NP-Hard problems – Kaminsky & Lloyd, MIT.
  http://arxiv.org/pdf/quant-ph/0211152
- The Ising model is NP-Complete, Barry A. Cipra. www.siam.org/pdf/news/654.pdf
- Test functions for optimization needs, Marcin Molga, Czesław Smutnicki.
  www.zsd.ict.pwr.wroc.pl/files/docs/functions.pdf
- Accelerated stochastic sampling of discrete statistical systems http://arxiv.org/abs/1010.0736
- Optimization by Quantum Annealing: Lessons from simple cases http://cdsweb.cern.ch/record/820862?ln=es
- Adiabatic Quantum Computation (et al.), Dorit Aharonov http://www.cs.huji.ac.il/~doria/papers.html
- Image recognition with an adiabatic quantum computer – I. Mapping to quadratic unconstrained binary optimization http://arxiv.org/abs/0804.4457
- Ground-state statistics from annealing algorithms: Quantum vs classical approaches
  http://www.citebase.org/abstract?id=oai%3AarXiv.org%3A0808.0365

- Adiabatic Quantum Computation is Equivalent to Standard Quantum Computation http://dl.acm.org/citation.cfm?id=1033159
- Adiabatic quantum computation fails for random instances of NP-complete problems http://arxiv.org/abs/0908.2782
- Quantum viscosity in a strongly interacting Fermi gas http://icamconferences.org/rpmbt15-09/documents/Thomas_OhioStatePrelimJuly09.pdf
- Quantum annealing: An introduction and new developments, Ohzeki & Nishimori http://cdsweb.cern.ch/record/1270899?ln=es
- Quantum Annealing for Variational Bayes inference: http://cdsweb.cern.ch/record/1178832?ln=es
- Adiabatic quantum computation, Enej Ilievski, University of Ljubljana http://mafija.fmf.uni-lj.si/seminar/files/2009_2010/adiabatic.pdf
- http://es.wikipedia.org/wiki/Hamiltoniano_(mec%C3%A1nica_cu%C3%A1ntica)
- Faster Annealing Schedules for Quantum Annealing, Satoshi Morita http://cdsweb.cern.ch/record/1023241?ln=es
- Code Generator for Quantum Simulated Annealing by Robert Tucci http://arxiv.org/abs/0908.1633
- QusAnn and Multiplexor expander Java applets http://www.ar-tiste.com/qusann.html
- The Quantum Annealing and its application in a classical computer http://www.smapip.is.tohoku.ac.jp/~smapip/2005/NHC+SMAPIP/ExtendedAbstracts/SeiSuzuki.pdf
- Hybrid quantum annealing for clustering problems: http://cdsweb.cern.ch/record/1345450?ln=es
- Energy gaps in quantum first-order mean-field-like transitions: The problems that quantum annealing can't solve, Jörg et al. http://arxiv.org/abs/0912.4865
- Ensemble equivalence in spin systems with short-range interactions, Takahashi, Nishimori y Martín-Mayor http://fts21.accesowok.fecyt.es/iopscience/1742-5468/2011/08/P08024?fromSearchPage=true
- Quantum annealing of an Ising spin-glass by Green's function Monte Carlo http://arxiv.org/abs/cond-mat/0608420
- Quantum thermal annealing with path integral Monte Carlo http://www.columbia.edu/cu/chemistry/groups/berne/papers/jpcA_104_86_2000.pdf
- Quantum annealing of a hard combinatorial problem, Daniel Lecina http://upcommons.upc.edu/pfc/handle/2099.1/11313

### Web and multimedia references:

- http://en.wikipedia.org/wiki/Hamiltonian_(quantum_mechanics)
- Quantum annealing (discussion pages) http://en.wikipedia.org/wiki/Talk:Quantum_annealing
- Sachiko Kodama's ferrofluids http://5magazine.wordpress.com/2011/05/31/sachiko-kodamas-ferrofluids/
- Absolute Astronomy | Carbon http://www.absoluteastronomy.com/topics/Carbon
- Helmholtz free energy http://en.wikipedia.org/wiki/Helmholtz_free_energy
- Diamante | Historia natural http://es.wikipedia.org/wiki/Diamante#Historia_natural
- Review Article Spin liquids in frustrated magnets http://www.nature.com/nature/journal/v464/n7286/full/nature08917.html
- Crystal structure http://en.wikipedia.org/wiki/Crystal_structure
- Néel temperature http://en.wikipedia.org/wiki/N%C3%A9el_temperature
- Spin-wave excitations in the ferromagnetic-metallic and in the charge, orbital and spin ordered states http://prb.aps.org/abstract/PRB/v84/i11/e094453
- Quantum Quenching, Annealing and Computation Anjan Kumar Kumar Chandra, Arnab Das and Bikas K. K. Chakrabarti http://www.springerlink.com/content/978-3-642-11469-4#section=742342&page=2&locus=18
- eNotes: Curse of dimensionality http://www.enotes.com/topic/Curse_of_dimensionality
- Quantum annealing and related optimization methods - Arnab Das, Bikas K. Chakrabarti http://books.google.es/books?id=m04GZnNyJ7MC&lpg=PA3&ots=WUY2iutMXb&dq=Quantum%20Annealing%20and%20Related%20Optimisation%20Methods&lr&hl=es&pg=PP1#v=twopage&q&f=false

- What is quantum tunneling? http://www.davidcolarusso.com/blog/?p=33#more-33
- El camino cuántico de dos fotones a lo largo de una línea recta http://francisthemulenews.wordpress.com/2010/09/20/el-camino-cuantico-de-dos-fotones-a-lo-largo-de-una-linea-recta/
- Simulaciones cuánticas, efecto túnel y doble rejilla en Java: Electron Technologies Models http://et.concord.org/software/
- Avances en el almacenamiento de datos http://avances-nanotecnologia.euroresidentes.com/2010/02/avances-en-el-almacenamiento-de-datos.html
- Fritipedia | Annealing http://www.frit-happens.co.uk/wiki/Annealing
- PBS Nova: The elegant universe http://www.pbs.org/wgbh/nova/physics/elegant-universe-einstein.html
- El Tamiz: cuántica sin fórmulas http://eltamiz.com/cuantica-sin-formulas/
- MathWorks example of Rastrigin's function http://www.mathworks.es/help/toolbox/gads/f14773.html
- Quantumaniac http://quantumaniac.tumblr.com/post/10373804268/quantum-tunneling-according-to-quantum-mechanics
- N-body problem integration displaying the actual Solar System http://www.lactamme.polytechnique.fr/Mosaic/images/NCOR.UA.16.D/display.html
- Gallery: Celestial Mechanics http://www.lactamme.polytechnique.fr/Mosaic/descripteurs/Galerie_CelestialMechanics.FV.html
- Plegamiento de proteínas: un problema interdisciplinario http://redalyc.uaemex.mx/src/inicio/ArtPdfRed.jsp?iCve=47548114
- From Levinthal to pathways to funnels http://www.dillgroup.ucsf.edu/danny/NatStructBiol/
- Is quantum annealing better than classical? http://cdsweb.cern.ch/record/569291?ln=es
- Adiabatic theorem http://en.wikipedia.org/wiki/Adiabatic_theorem
- Spin ice http://en.wikipedia.org/wiki/Spin_ice
- Zero-point energy http://en.wikipedia.org/wiki/Zero-point_energy
- Geometrically frustrated magnetism http://www.phys.psu.edu/~schiffer/research/
- Efectos Cuánticos I: La coherencia http://entangledapples.blogspot.com/2010/11/efectos-cuanticos-i-la-coherencia.html
- Coherent states http://en.wikipedia.org/wiki/Coherent_state
- Shape-Invariant Solutions of the Quantum Fokker-Planck Equation for an Optical Oscillator http://demonstrations.wolfram.com/ShapeInvariantSolutionsOfTheQuantumFokkerPlanckEquationForAn/
- The Causal Interpretation of Quantum Tunneling through a Square Barrier and Well http://demonstrations.wolfram.com/TheCausalInterpretationOfQuantumTunnelingThroughASquareBarri/
- Computación cuántica: teoría y algoritmos http://entangledapples.blogspot.com/2011/06/computacion-cuantica-teoria-y.html
- Espacio de Hilbert http://es.wikipedia.org/wiki/Espacio_de_hilbert
- D-Wave Systems sells its first quantum computing system lo Lockheed Martin corp. http://www.dwavesys.com/en/pressreleases.html#lm_2011
- Quantum Annealing can be millions of times faster than classical computing http://nextbigfuture.com/2008/02/quantum-annealing-millions-of-times.html
- Dwave Systems announces 512 qubit Adiabatic Quantum Computer before end of 2012 http://nextbigfuture.com/2011/11/dwave-systems-announces-512-qubit.html
- Dwave Systems official blog http://dwave.wordpress.com
- Francis (th)E mule Science's News | Computación cuántica http://francisthemulenews.wordpress.com/tag/computacion-cuantica/
- Quantum annealing with manufactured spins – Nature http://www.nature.com/nature/journal/v473/n7346/full/nature10012.html
- Google Tech Talks on D-Wave and Quantum Annealing http://dwave.wordpress.com/2010/11/01/google-tech-talks-on-d-wave-and-quantum-annealing/
- Quantum Computing Day (1 & 2) Google Tech Talks http://www.youtube.com/watch?v=I56UugZ_8DI
- Google Workshop on Quantum Biology, "Learning from Examples Using Quantum Annealing" Presented by

Hartmut Neven, October 22, 2010 http://www.youtube.com/watch?v=HKUZ6IuJyHw

- Machine Learning in Google Googles http://techtalks.tv/talks/54457/
- Training a binary classifier with the quantum adiabatic algorithm http://videolectures.net/opt08_neven_tabcwt/
- Who needs doors when I can tunnel? - Javier Rodríguez Laguna http://physicsnapkins.wordpress.com/2011/05/13/who-needs-doors-when-i-can-tunnel/
- QA in m3l http://code.google.com/p/m3l/source/detail?r=128&path=/trunk/options.c#
- Path Integral Monte Carlo in C++ http://cms.mcc.uiuc.edu/pimcpp/index.php/Main_Page
- Quantum Annealing of the graph coloring problem – Olawale Titiloye, Alan Crispin
- QuMax library http://www.attaccalite.altervista.org/qumax/index.php
- QuMax at SourceForge http://sourceforge.net/projects/qumax/

**Author works:**

- http://es.scribd.com/doc/66288537/Temple-paralelo-Estado-del-arte
- http://alfonsoycia.blogspot.com/2011/05/visitando-en-cern-lhc-atlas-en-ginebra.html
- http://alfonsoycia.blogspot.com/2010/04/cortoplacismo-y-el-principio-de.html
- http://alfonsoycia.blogspot.com/2007/07/matemtica-cuntica-matemtica.html
- http://alfonsoycia.blogspot.com/2010/02/simulacion-de-ondas-en-3d.html